\documentclass[pre,aps,showkeys,twocolumn]{revtex4}
\usepackage{graphicx}
\usepackage{dcolumn}
\usepackage{bm}
\usepackage{amsmath,amssymb}
\usepackage{color,soul}

\begin{document}

\title{Conventional and rotating magnetoelectric effect of a half-filled spin-electron model on a doubly decorated square lattice}
\author{Hana \v Cen\v carikov\'a$^{1}$ and Jozef Stre\v{c}ka$^2$}
\affiliation{
$^1$Institute of Experimental Physics, Slovak Academy of Sciences, Watsonova 47, 040 01 Ko\v {s}ice, Slovakia \\ 
$^2$ Department of Theoretical Physics and Astrophysics, Faculty of Science, P. J. \v{S}af\'{a}rik University, Park 
Angelinum 9, 040 01 Ko\v{s}ice, Slovakia}

\begin{abstract}
 A conventional and rotating magnetoelectric effect of a half-filled spin-electron model on a doubly decorated square lattice is investigated by exact calculations.
An importance of the electron hopping and spatial orientation of the electric field upon a magnetoelectric effect is examined in detail. A critical temperature may display one or two consecutive round maxima as a function of the electric field. Although the rotating magnetoelectric effect (RME) does not affect the ground-state ordering, the pronounced RME is found close to a critical temperature of continuous phase transition. It is shown that RME is amplified upon strengthening of the electric field, which additionally supports thermal fluctuations in destroying a spontaneous antiferromagnetic long-range order.
\end{abstract}

%\pacs{75.10.Hk, 75.85.+t, 68.35.Rh, 75.10.−b}
\keywords{Classical spin models; Magnetoelectric effects; Phase transition and critical phenomena; General theory and models of magnetic ordering}
\maketitle
\section{Introduction}\label{sec:introduction}
More than 3000 publications dealing with a magnetoelectric effect registered in research databases (Web of Science and Scopus) during last five years document a huge attractiveness of unconventional materials, whose magnetic properties can be controlled by an electric field. It is apparent that a wide range of practical applications e.g., in  the spintronics, automation engineering, security, navigation or medicine~\cite{Prinz,Wolf,Son,Scott,Hur,Wu,Vopson,Sreenivasulu}, is a main reason of such expanded activities. Another driving force stimulating an investigation of magnetoelectric materials lies in their potential to create    novel smart low-energy consuming and/or eco-friendly devices, which will be a more regardful to an environment. Unfortunately, a relatively small response  of aforementioned materials with respect to an applied electric or magnetic field, represents a serious technological deficiency (see Refs.~\cite{Fiebig,Tokura} and references cited therein). Consequently, an  alternative mechanism improving a magnetoelectric response is highly desirable. Taking into account the phenomenological theory of a magnetoelectric effect developed by Dzyaloshinskii~\cite{Dzyalosh}, in which its existence is conditioned by an antisymmetric anisotropy of an investigated system, affords a possible alternative way enhancing response of a magnetic system on an applied electric field  achieved by its spatial rotation. In a consequence of that, the charged particles from different spatial directions are subject to a different electric field, which may thus strengthen magnetic anisotropy due to a magnetoelectric effect.  A similar parallelism can be found in a rotating magnetocaloric effect~\cite{Nikitin}, where the rotary magnetic refrigeration turns out to be technically more convenient and effective in comparison to its conventional counterpart~\cite{Zhang,Caro,Lorusso,Balli1,Balli3,Orendac,Moon}.  

In the present work, we will rigorously examine  a conventional and rotating magnetoelectric effect in a coupled spin-electron model on a doubly decorated square lattice introduced in Ref.~\cite{Doria,Cenci2015} in the physically most interesting half-filled case. This hybrid two-dimensional (2D) model exhibits  various unconventional cooperative phenomena, like a magnetic reentrance~\cite{Cenci2016}, metamagnetic transitions~\cite{Cenci2018a} or enhanced magnetoelectric effect~\cite{Cenci2018}, and its relative simplicity makes it solvable by using rigorous methods. It should be pointed out that there are only a few rigorous studies focusing on the magnetoelectric effect of one-dimensional quantum spin chains~\cite{Brockmann,Menchyshyn,Baran}, zero-dimensional Hubbard pair~\cite{Balcerzak1,Balcerzak2,Balcerzak3} or cubic clusters~\cite{Szalowski}, which even makes the introduced model a more attractive for further investigation. 
 To the best of our knowledge, the magnetoelectric effect of a magnetic system either placed in a spatially rotating electric field or more conventionally the magnetoelectric effect of a magnetic system rotating in a static (constant) electric field has not been investigated yet. In the spirit of these facts, we hope that our analyses will bring a new insight into characteristic features of the rotating magnetoelectric effect and will be stimulating for further research activities in this field of study.

This Letter is organized as follows. The model under the investigation is briefly introduced in Sec.~\ref{model} along with basic steps of its analytical treatment. The most interesting results demonstrating existence of a magnetoelectric effect are discussed in Sec.~\ref{results}. Finally, several concluding remarks are mentioned in Sec.~\ref{conclusions}.

\section{Model and Method}
\label{model}
\begin{figure}[h!]
{\includegraphics[width=.2\textwidth,height=3.7cm,trim=1.9cm 5cm 8.9cm 12cm, clip]{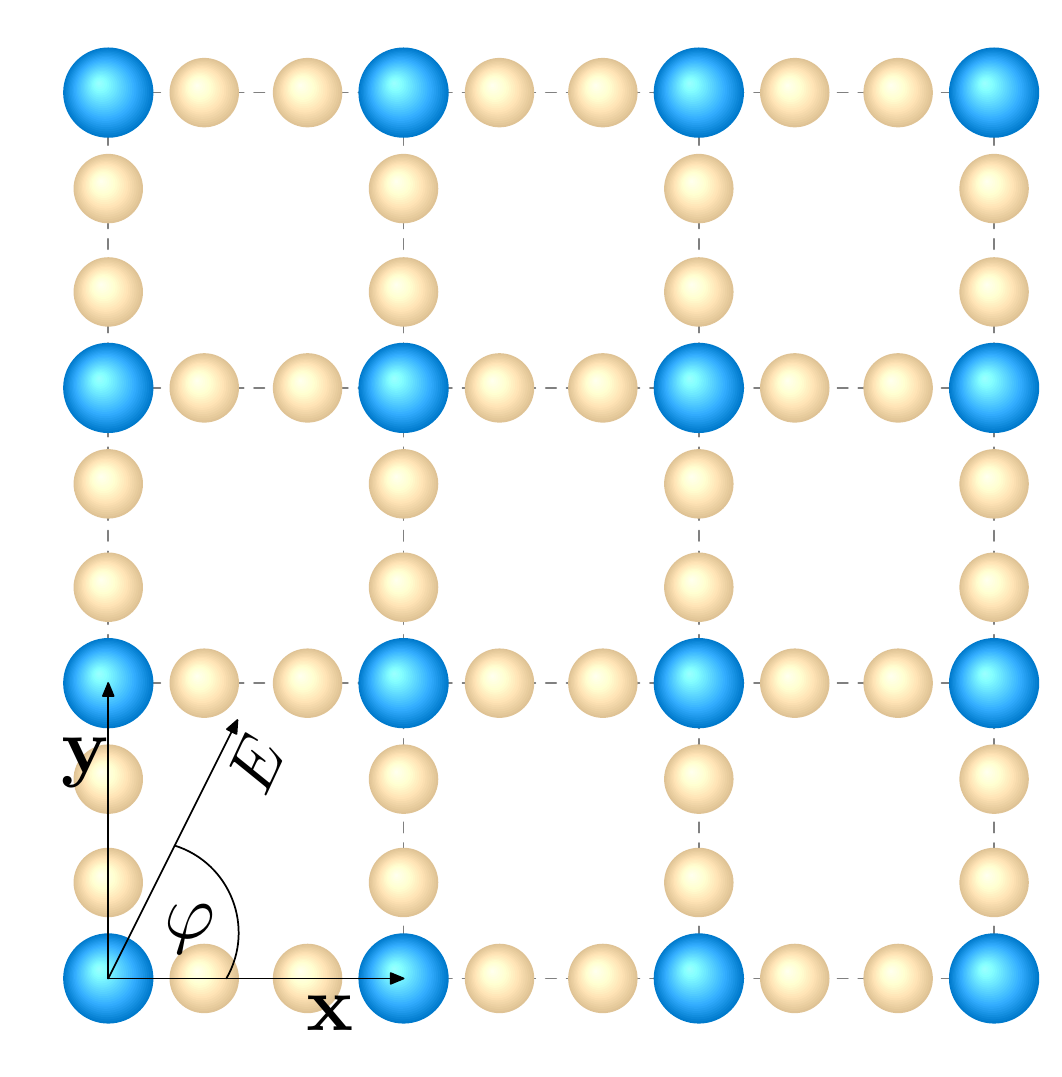}}\hspace*{0.5cm}
{\includegraphics[width=.2\textwidth,height=3.7cm,trim=2.8cm 6.5cm 12.3cm 14.5cm, clip]{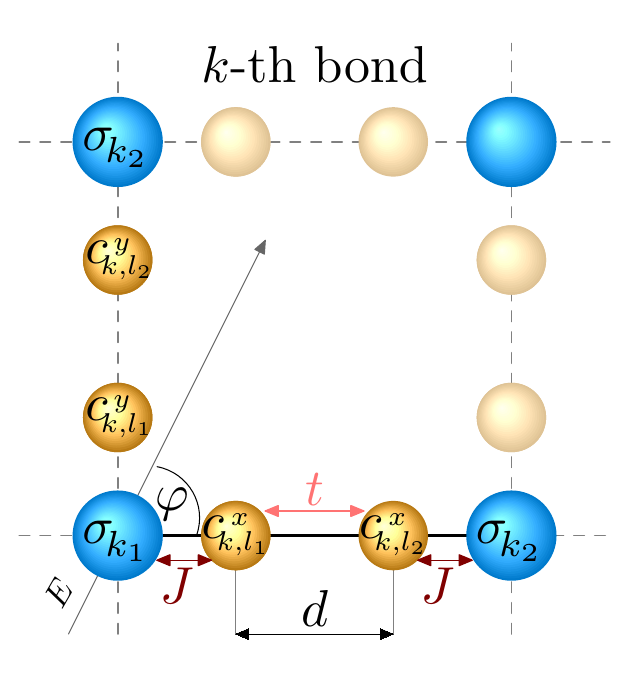}}
\caption{ A schematic representation of a doubly decorated square lattice (left panel) with a detail of a selected $k$-th horizontal and vertical bond (right panel). Large (blue) balls illustrate location of the Ising spins and small (yellow) balls illustrate position of the decorating sites over which the mobile electrons are delocalized. All assumed interactions within the investigated model are also displayed.}  
\label{fig0}
\end{figure}

\begin{figure}[h!]
{\includegraphics[width=.35\textwidth]{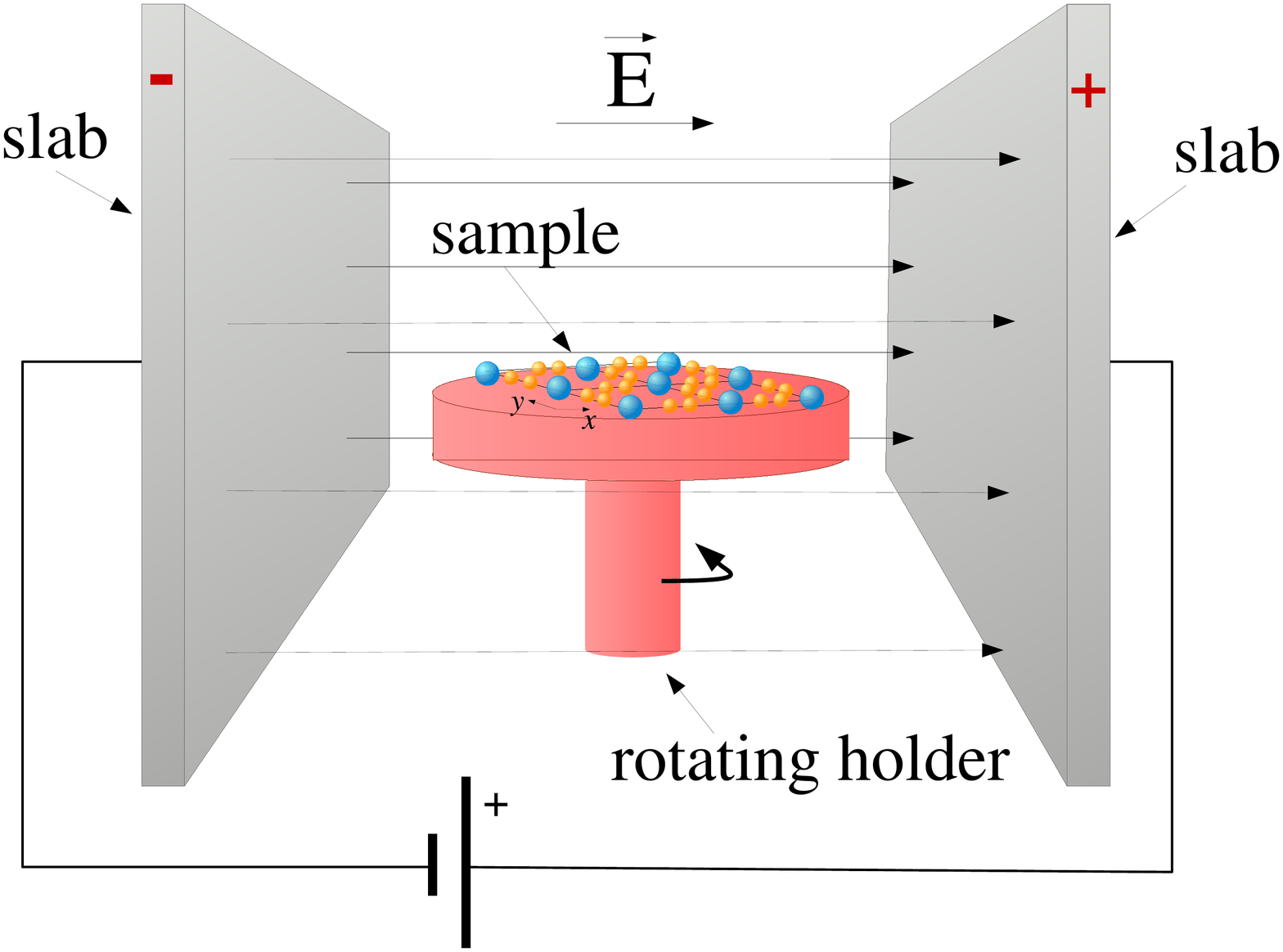}}
\caption{ A set-up for an investigation of the rotating magnetoelectric effect when the sample rotates in a homogeneous electric field of constant magnitude.} 
\label{setup}
\end{figure}

Let us consider a hybrid spin-electron model on a doubly decorated square lattice, which consists of localized Ising spins and a set of mobile electrons placed at two decorating lattice sites positioned between two nearest-neighbor Ising spins  as displayed in Fig.~\ref{fig0}. 
Our interest is focused just on a special half-filled case, when considering exactly two mobile electrons per each couple of the decorating sites (bond). The relevant interactions of the model under investigation determining the magnetic and electric properties relate to a quantum-mechanical motion of the mobile electrons delocalized over two nearest-neighbor decorating sites characterized by the hopping term $t$, the exchange interaction $J$ between the localized Ising spin and its nearest-neighbor mobile electron(s), as well as, the electrostatic potential energy $V$ of the mobile electrons. The total Hamiltonian $\hat{\cal H}$ can be expressed as a sum of $N$ bond Hamiltonians referring to a horizontal direction ($\hat{\cal H}_k^x$) and $N$ bond Hamiltonians referring to a vertical direction ($\hat{\cal H}_k^y$)  
\allowdisplaybreaks
\begin{eqnarray}
\hat{\cal H}\!=\!\sum_{k=1}^N\hat{\cal H}_k^x\!+\!\sum_{k=1}^N\hat{\cal H}_k^y,
\label{eq1}
\end{eqnarray} 
 whereas both bond Hamiltonians $\hat{\cal H}_k^{\delta}$ ($\delta\!=\!x,y$) can be uniquely defined as follows 
\allowdisplaybreaks
\begin{eqnarray}
\hat{\cal H}^{\delta}_k=\!\!\!&-&\!\!\!t(\hat{c}^{\dagger,\delta}_{k,l_1,\uparrow}\hat{c}^{\delta}_{k,l_2,\uparrow}\!+\!\hat{c}^{\dagger,\delta}_{k,l_1,\downarrow}\hat{c}^{\delta}_{k,l_2,\downarrow}\!+\! H.c.)
\nonumber
\\
\!\!\!&-&\!\!\!J \hat{\sigma}^{z}_{k_1}({n}^{\delta}_{k,l_1,\uparrow}-\hat{n}^{\delta}_{k,l_1,\downarrow})\!-\!
J \hat{\sigma}^{z}_{k_2}({n}^{\delta}_{k,l_2,\uparrow}\!-\!\hat{n}^{\delta}_{k,l_2,\downarrow})
\nonumber\\
\!\!\!&-&\!\!\!V_{\delta}(\hat{n}^{\delta}_{k,l_1,\uparrow}\!+\!\hat{n}^{\delta}_{k,l_1,\downarrow})\!+\!
V_{\delta}(\hat{n}^{\delta}_{k,l_2,\uparrow}\!+\!\hat{n}^{\delta}_{k,l_2,\downarrow}).
\label{eq2a}
\end{eqnarray}
Here, $\hat{c}^{\dagger,\delta}_{k,l_{\alpha},\gamma}$ and $\hat{c}^{\delta}_{k,l_{\alpha},\gamma}$ ($\alpha\!=\!1,2$, $\gamma\!=\!\uparrow,\downarrow$ and $\delta\!=\!x,y$) denote the creation and annihilation Fermi operators of the mobile electrons from the $k$-th bond with the respective number operator $\hat{n}_{k,l_{\alpha},\gamma}^{\delta}\!=\!\hat{c}^{\dagger,\delta}_{k,l_{\alpha},\gamma}\hat{c}^{\delta}_{k,l_{\alpha},\gamma}$. Symbols ${\hat\sigma}^{z}_{k_{\alpha}}$ are the standard  spin operators with eigenvalues 
${\sigma}^{z}_{k_{\alpha}} = \pm 1$. It is worthwhile to recall that the influence of an electric field on a subsystem of the mobile electrons is contained in the electrostatic potential energy $V_{x(y)}\!=\!E_{x(y)}|e|d/2$, whose magnitude is unambiguously specified by the intensity of an external electric field $E$ and a distance $d$ of two nearest-neighbor decorating sites comprising the mobile electrons with a charge $|e|$.  Although the electric-field terms involved in the bond Hamiltonians (\ref{eq2a}) are somewhat reminiscent of the chemical potential of the mobile electrons, they generally prescribe different electrostatic energy for two decorating sites on all horizontal as well as all vertical bonds depending on the polar angle $\varphi$. Note that the polar angle $\varphi\in(0,\pi/2)$ unambiguously determines a spatial orientation of the external electric field ranging from the crystallographic axis \textbf{[10]} to \textbf{[01]}. Owing to this fact, an applied electric field may discriminate either left or right (upper or lower) decorating site from the horizontal (vertical) bonds with respect to the occupancy by the mobile electrons. For further convenience, the aforementioned differences of the electrostatic potential energy of the bond Hamiltonians (\ref{eq2a}) of the horizontal and vertical bonds can be mathematically parametrized through its magnitude $V$ and polar angle $\varphi$ as $V_x\!=\!V \cos \varphi$ and $V_y\!=\!V \sin \varphi$ when taking into account two orthogonal projections of the external electric field under a square-lattice geometry. In addition to the conventional magnetoelectric effect investigating response of a magnetic system with respect to the external electric field of varying size under the imposed spatial orientation $\varphi$, we may also alternatively investigate the rotating magnetoelectric effect for a magnetic system rotating in the external electric field of constant magnitude. A set-up for an investigation of the rotating magnetoelectric effect is schematically illustrated in Fig. \ref{setup} for the particular case when the sample rotates in a homogeneous electric field of constant magnitude.

The thermal analyses of the model under investigation can be performed rigorously using the generalized mapping transformations~\cite{Fisher,Syozi,Rojas,Strecka2010}. 
In this procedure the considered spin-electron model can alternatively be perceived as a pure localized Ising model with novel effective parameters. As a result of a straightforward algebra comprehensively described in our previous works~\cite{Cenci2018,Cenci2019}, one may obtain a simple relation between the canonical partition function ${\cal Z}$ of an investigated model and a canonical partition function of a pure spin-1/2 Ising model  ${\cal Z}_{IM}$ on a rectangular lattice
\begin{eqnarray}
{\cal Z}(\beta,t,J,V,\varphi)\!=\!(A^{x})^N(A^{y})^N{\cal Z}_{IM}(\beta,R^{x},R^{y}),
\label{eq01}
\end{eqnarray}
where $\beta\!=\!1/(k_BT)$ denotes the reciprocal temperature normalized by the inverse Boltzmann's constant $k_B$ and $T$ is an absolute temperature. The explicit expressions of mapping parameters $A^{\delta}$, $\beta R^{\delta}$ may be subsequently obtained from the self-consistent condition
\begin{eqnarray}
{\cal Z}_k^{\delta}(\beta,t,J,V_{\delta})\!=\!A^{\delta}\exp(\beta R^{\delta}\sigma_{k_1}^z\sigma_{k_2}^z),
\label{eq03}
\end{eqnarray}
which should hold for any combination of the Ising spins $\sigma_{k_1}^z$ and $\sigma_{k_2}^z$. From this set of equations one gets
\begin{eqnarray}
A^{\delta}\!=\!(V^{\delta}_1V^{\delta}_2)^{1/2}, \beta R^{\delta}\!=\!\frac{1}{2}\ln\left(\frac{V^{\delta}_1}{V^{\delta}_2}\right),
\label{eq05}
\end{eqnarray}
where the functions $V_1^{\delta}$ and $V_2^{\delta}$ are defined as follows
\begin{eqnarray}
V^{\delta}_1\!\!\!&=&\!\!\!2\left(1\!+\!\cosh2\beta J\!+\!\cosh2\beta\sqrt{(V_{\delta})^2\!+\!t^2}\right),\nonumber\\
V^{\delta}_2\!\!\!&=&\!\!\!2\left(1+\cosh\beta f_{+}^{\delta}\!+\!\cosh\beta f_{-}^{\delta}\right),\nonumber\\
f^{\delta}_{\pm}\!\!\!&=&\!\!\!\sqrt{2(J^2\!+\!(V_{\delta})^2\!+\!t^2)\pm2b^{\delta}_{+}b^{\delta}_{-}},\nonumber\\
b_{\pm}^{\delta}\!\!\!&=&\!\!\!\sqrt{(J\pm2V_{\delta})^2\!+\!t^2}\;.
\label{eq06}
\end{eqnarray}

In order to analyse the critical behaviour of an investigated model, we can determine the critical temperature from the following condition 
\begin{eqnarray}
\sinh(2\beta_c R^x)\sinh(2\beta_c R^y)\!=\!1,
\label{eq07}
\end{eqnarray}
which is consistent with Onsager's critical condition derived for the spin-1/2 Ising model on a rectangular lattice~\cite{Onsager}. We note that the lower index $c$ in the critical condition~(\ref{eq07}), naturally, points out that the inverse critical temperature $\beta_c$ enters into this relation. 

 For a completeness, let us define the staggered magnetization of the Ising spins ($m_i^{\delta}$) 
\begin{eqnarray}
m_i^{\delta}\!\!\!&=&\!\!\!\frac{1}{2}\langle \hat{\sigma}^z_{k_1}\!-\!\hat{\sigma}^z_{k_2}\rangle
\nonumber\\
\!\!\!&=&\!\!\!\left[1-\sinh^{-2}(2|\beta R^x|)\sinh^{-2}(2|\beta R^y|)\right]^{1/8},
\label{eq12}
\end{eqnarray}
which follows from the well-know formula originally derived by Potts~\cite{Potts} and Chang~\cite{Chang} for the spin-1/2 Ising model on the rectangular lattice.
The staggered magnetization of the mobile electrons ($m_e^{\delta}$) can be calculated according to the formula
\begin{eqnarray}
m_e^{\delta}\!\!\!&=&\!\!\!\left\langle \sum_{\alpha=1}^2\frac{(-1)^{\alpha}}{2}\left(\hat{n}_{k,l_{\alpha,\downarrow}}\!-\!\hat{n}_{k,l_{\alpha,\uparrow}}\right)\right\rangle
\nonumber\\
\!\!\!&=&\!\!\!\left\langle - \frac{1}{2{\cal Z}_k^{\delta}}\sum_{\alpha=1}^2
\frac{(-1)^{\alpha}\partial {\cal Z}_k^{\delta}}{\partial \beta J\sigma_{k_{\alpha}}}
\right\rangle
\!=\!
\left(\left(1\!+\!\Lambda_{\delta}\right)\frac{\sinh\beta f_{+}^{\delta}}{f_{+}^{\delta}}\right.
\nonumber\\
\!\!\!&+&\!\!\!\left.\left(1\!-\!\Lambda_{\delta}\right)\frac{\sinh\beta f_{-}^{\delta}}{f_{-}^{\delta}}\right)
\frac{Jm_i^{\delta}}{1\!+\!\cosh\beta f_{+}^{\delta}\!+\!\cosh\beta f_{-}^{\delta}},
\label{eq13}
\end{eqnarray}
where $\Lambda_{\delta}\!=\!\frac{J^2\!-\!V^2_{\delta}\!+\!t^2}{b_{+}^{\delta}b_{-}^{\delta}}$.

\section{Results and discussion}
\label{results}
Before proceeding a discussion of the most interesting results, it is noteworthy to mention that all calculations will be hereafter performed for a ferromagnetic interaction $J$ ($J\!>\!0$), since its substitution with an antiferromagnetic counterpart causes only a trivial interchange of a relative spin orientation and has not a substantial impact on phase diagrams and thermodynamics. Without loss of generality, all values of the model parameters will be normalized with respect to the exchange coupling $J$. In addition, we may also apply a limitation for a polar angle $\varphi$ ranging from $\varphi\!=\!0$ to $\varphi\!=\!\pi/4$, since the complementary polar angle $\varphi\!\in\!(\pi/4,\pi/2)$ stabilizes identical magnetic structures rotated about the angle of $\pi/2$.

\begin{figure}[h!]
{\includegraphics[width=.4\textwidth,height=5.5cm,trim=3.cm 9.3cm 4cm 9.8cm, clip]{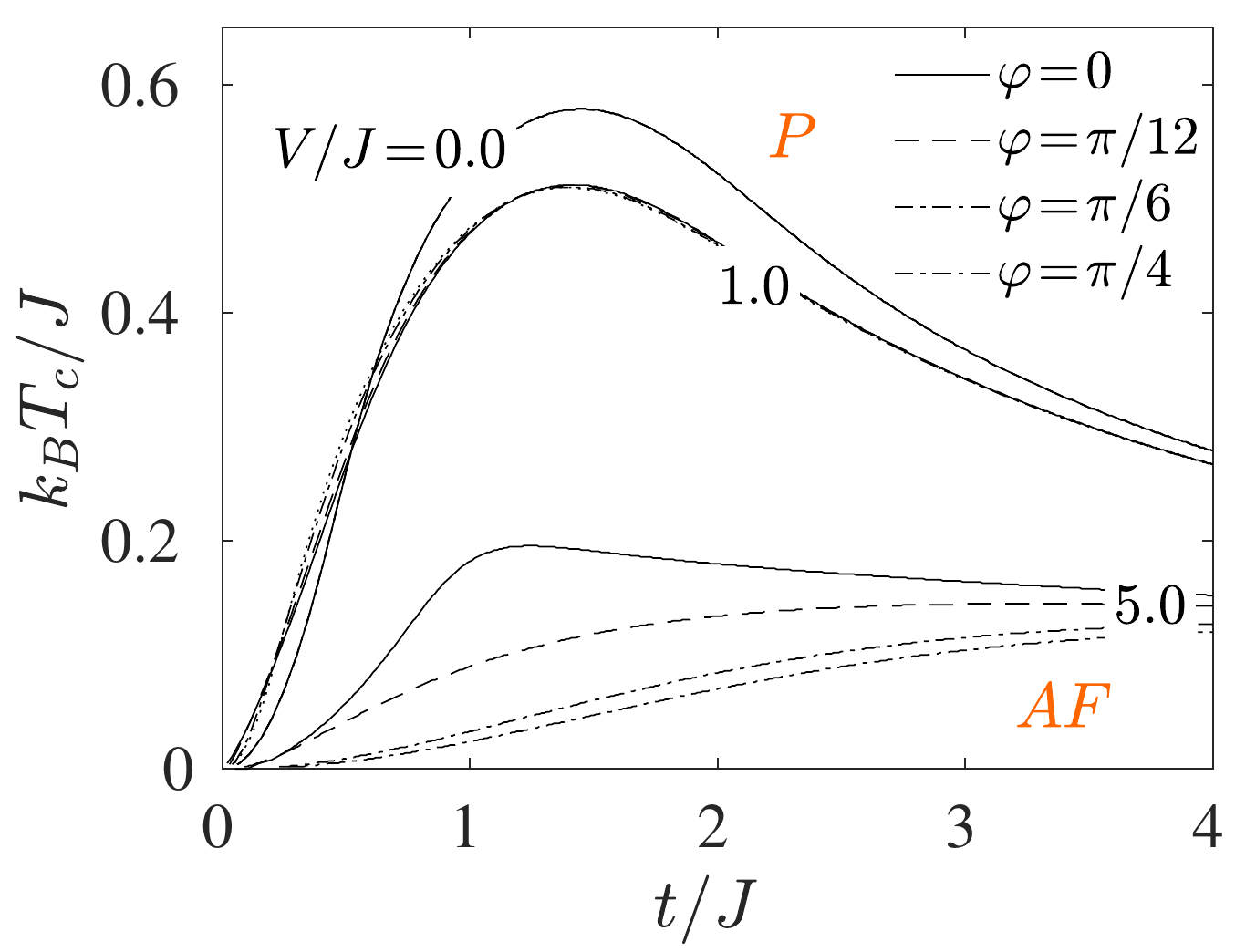}}
\caption{The normalized critical temperature $k_BT_c/J$ as a function of the hopping term $t/J$ for selected values of the electric field $V/J$ and polar angle $\varphi$.}
\label{fig2}
\end{figure}

\begin{figure}[h!]
{\includegraphics[width=.4\textwidth,height=5.5cm,trim=3cm 9.3cm 4cm 9.8cm, clip]{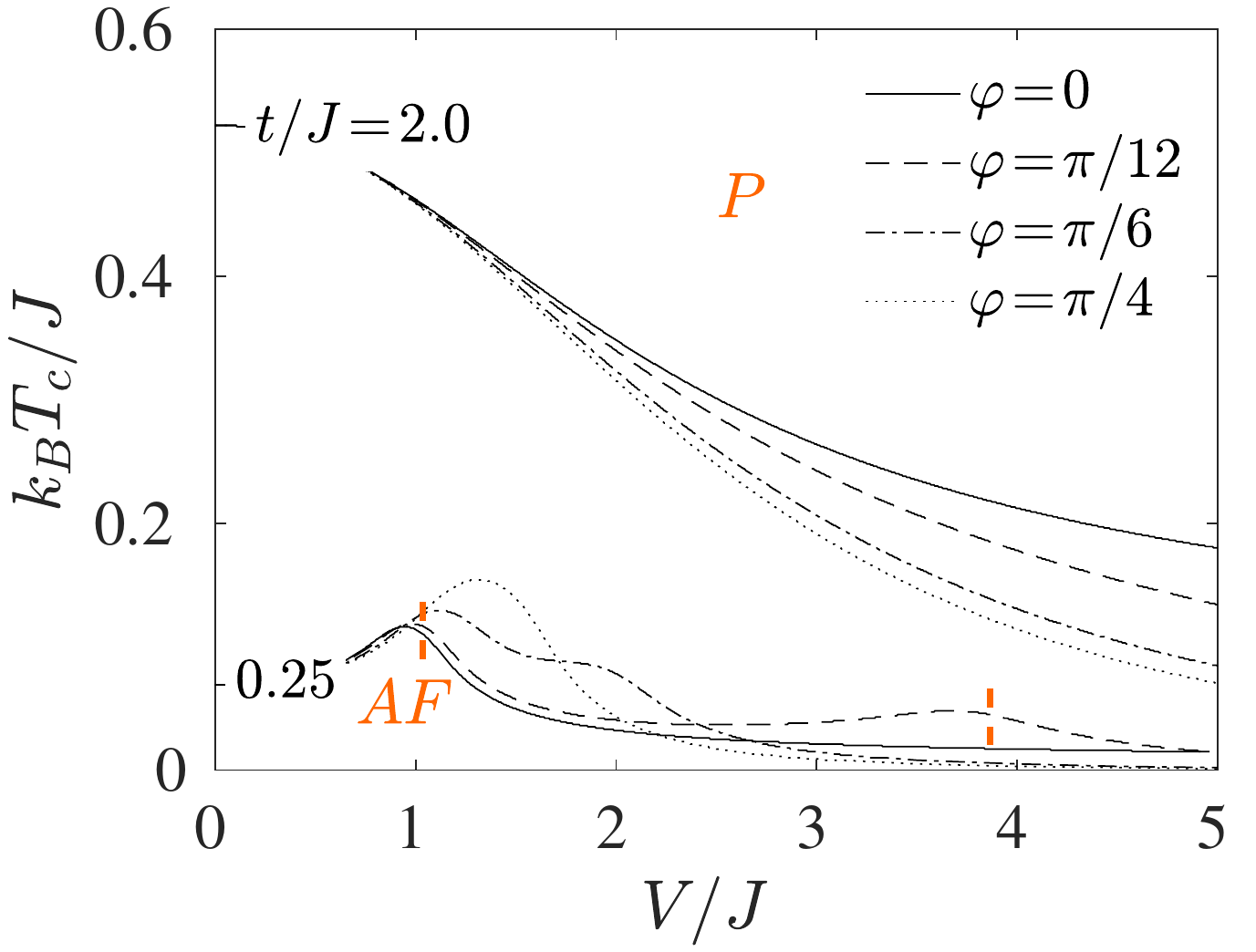}}
{\includegraphics[width=.4\textwidth,height=5.5cm,trim=3cm 9.3cm 4cm 9.8cm, clip]{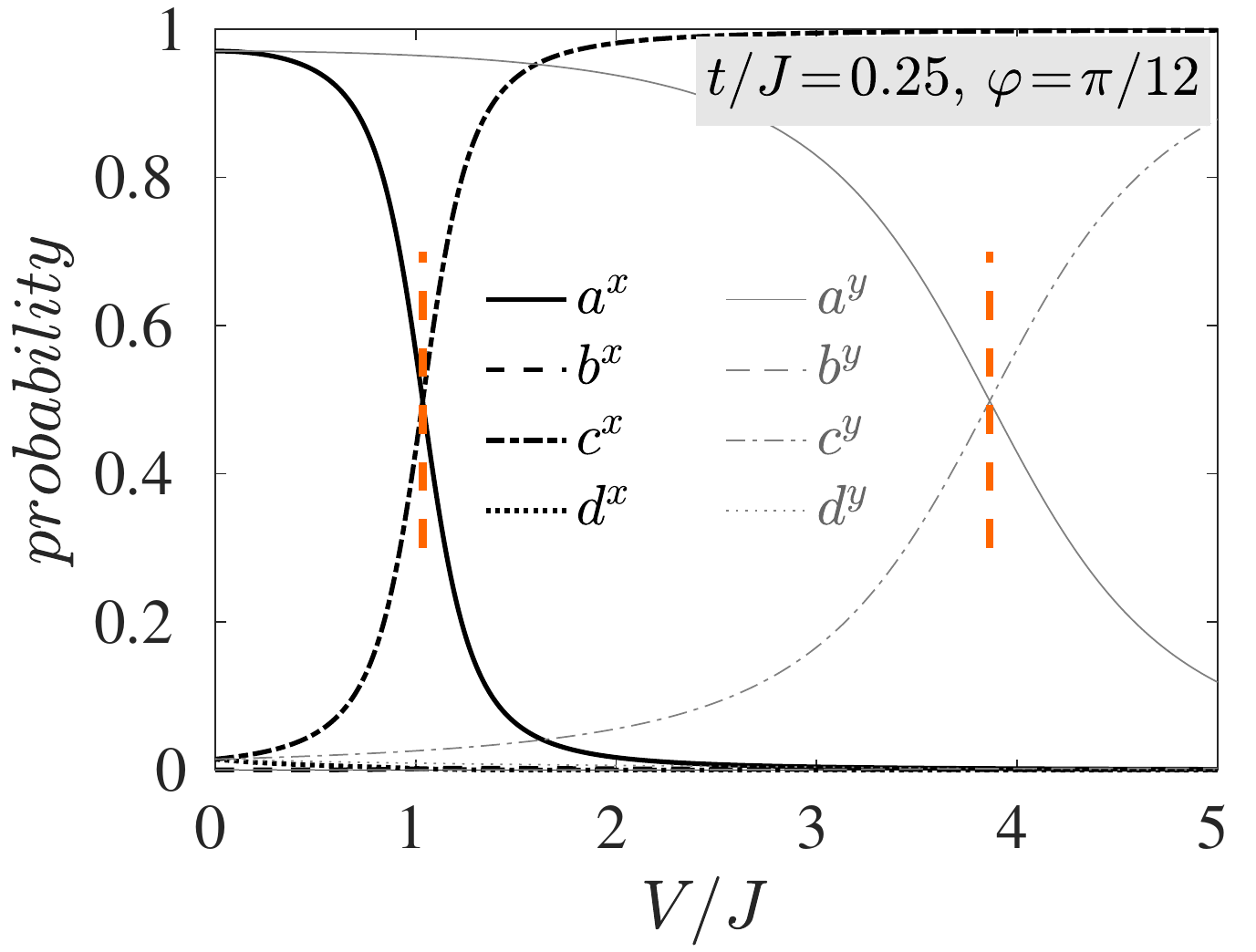}}
\caption{Upper part: The normalized critical temperature $k_BT_c/J$ as a function of the electric field $V/J$ for selected values of the hopping term $t/J$ and the polar angle $\varphi$. Lower part: The probability distribution for four entangled microstates (\ref{eq3b}) as a function of the electric field $V/J$ for $t/J\!=\!0.25$ and $\varphi\!=\!\pi/12$.}
\label{fig3}
\end{figure}

First we analyse the ground-state properties with an aim to determine all energetically favourable magnetic arrangements. Using a direct diagonalization of the bond Hamiltonians (\ref{eq2a}) one may identify just a single magnetic ground state $|{\rm AF} \rangle = \prod_{k=1}^N|{\rm II}\rangle^{x}_k|{\rm II}\rangle^{y}_k$ with character of the quantum antiferromagnetic phase, which is expressed in terms of the following eigenvectors of the horizontal and vertical bonds ($\delta\!=\!x,y$) 
\begin{eqnarray}
|\textrm{II}\rangle^{\delta}_k\!\!\!&=&\!\!\!|1\rangle_{\sigma_{k_1}}\otimes\left(a^{\delta}|\uparrow,\downarrow\rangle_k\!+\!b^{\delta}|\downarrow,\uparrow\rangle_k\!+\!c^{\delta}|\uparrow\downarrow,0\rangle_k\right.
\nonumber\\
\!\!\!&+&\!\!\!\left.d^{\delta}|0,\uparrow\downarrow\rangle_k\right)\otimes|-1\rangle_{\sigma_{k_2}},
\label{eq3a}\\
\nonumber\\
&&\hspace{-0.5cm}a^{\delta}=\frac{2t\xi_{\delta}}{\eta_{\delta}}(\xi_{\delta}\!+\!2J),\;\;
b^{\delta}=-a^{\delta}\frac{(\xi_{\delta}\!-\!2J)}{(\xi_{\delta}\!+\!2J)},
\nonumber\\
&&\hspace{-0.5cm}c^{\delta}=\frac{a^{\delta}}{2t\xi_{\delta}}\frac{(\xi_{\delta}\!-\!2J)}{(\xi_{\delta}\!+\!2V_{\delta})},\;\;
d^{\delta}=c^{\delta}\frac{(\xi_{\delta}\!-\!2V_{\delta})}{(\xi_{\delta}\!+\!2V_{\delta})},
\nonumber\\
&&\hspace{-1.cm}\xi_{\delta}\!=\!\left\{2\left[\left(J^2\!+\!V_{\delta}^2\!+\!t^2\right)\!+\! \frac{\left((J\!+\!V_{\delta})^2\!+\!t^2\right)^{1/2}}{\left((J\!-\!V_{\delta})^2\!+\!t^2\right)^{-1/2}}\right]\right\}^{1/2},
\nonumber\\
&&\hspace{-1.cm}\eta_{\delta}\!=\!\left\{2\left[4t^2{\xi_{\delta}}^2({\xi_{\delta}}^2\!+\!4J^2)
\!+\!\frac{({\xi_{\delta}}^2\!-\!4J^2)}{({\xi_{\delta}}^2\!+\!4V_{\delta}^2)^{-2}}\right]\right\}^{1/2}.
\label{eq3b}
\end{eqnarray}
Surprisingly, the antiferromagnetic ground state $|{\rm AF} \rangle$ is stable over the whole parameter space regardless of a size and a spatial orientation of the external electric field. However, the variation of an electric field (either its strength and/or its spatial orientation) diversely influences the mobile electrons on horizontal and vertical bonds, which could imply a possible electric-field induced anisotropy of magnetic and thermodynamic properties. From this point of view, the critical temperature $k_BT_c/J$ has been analysed in detail as a function of the size and spatial orientation of the electric field. In accordance with our expectations the critical temperature for an arbitrary polar angle $\varphi$  is almost identical at weak electric fields ($V/J\!\lesssim\!1$), but it is dramatically changed upon increasing of its relative size $V/J$. This critical behaviour can be seen in Figs.~\ref{fig2} and~\ref{fig3}.  
Another interesting observation, which follows from Fig.~\ref{fig2}, is that the increasing polar angle $\varphi$ can transiently stabilize the spontaneous antiferromagnetic ordering if both the applied electric field $V/J$ and the hopping integral $t/J$ are weak enough. By contrast, the increasing polar angle $\varphi$ generally reduces the critical temperature at sufficiently strong electric fields $V/J$. The possible explanation of these diverse scenarios could be found through the changes of occurrence probabilities of the intrinsically antiferromagnetic and non-magnetic microstates of the mobile electrons as derived from  Eq.~(\ref{eq3b}). In a strong electric field the increasing polar angle $\varphi$ substantially enhances (reduces) the occurrence probability for non-magnetic (antiferromagnetic) microstates of the mobile electrons (see lower part in Fig. \ref{fig3}), which results in lowering of the critical temperature regardless of the hopping term $t/J$. Contrary to this, the competition of all relevant interactions in a weak electric field leads to an increment of the occurrence probability of the antiferromagnetic microstates of the mobile electrons (the increment of $k_BT_c/J$) at weaker hopping amplitudes $t/J$, whereas the reverse trend, an enhancement of the occurrence probabilities of non-magnetic microstates (the decrement of $k_BT_c/J$), is detectable at higher hopping amplitudes $t/J$. 

Taking into account a sufficiently large electron hopping $t/J\!\gtrsim\!1.1$, the applied electric field gradually destabilizes a spontaneous antiferromagnetic long-range order 
and is responsible for a decreasing profile of the critical temperature demonstrated in Fig.~\ref{fig3}.  While the critical temperature vs. electric field plots depicted 
in Fig.~\ref{fig3} for the fixed value of the polar angle $\varphi$ serve in evidence of a conventional magnetoelectric effect, the variation of the critical temperature at the constant electric field achieved purely due to a change of its spatial orientation (i.e. a change of the polar angle $\varphi$) exemplifies a rotating magnetoelectric effect. In accordance with our previous discussion, special spatial orientation of the external electric field with $\varphi\!\to\!\pi/4$ lowering a system anisotropy supports thermal fluctuations when shifting the critical temperature to lower values. A marked difference of the critical temperature at high electric fields indicates a substantial rotating magnetoelectric effect on assumption that the hopping integral $t/J$ is sufficiently strong. Inspecting a weak hopping amplitude $t/J$ leads to a very intriguing behaviour of the critical temperature with one or two round maxima (see the case $t/J\!=\!0.25$ in Fig.~\ref{fig3}). 
\begin{figure}[b!]
{\includegraphics[width=.4\textwidth,height=5.5cm,trim=3cm 9.3cm 4.cm 9.8cm, clip]{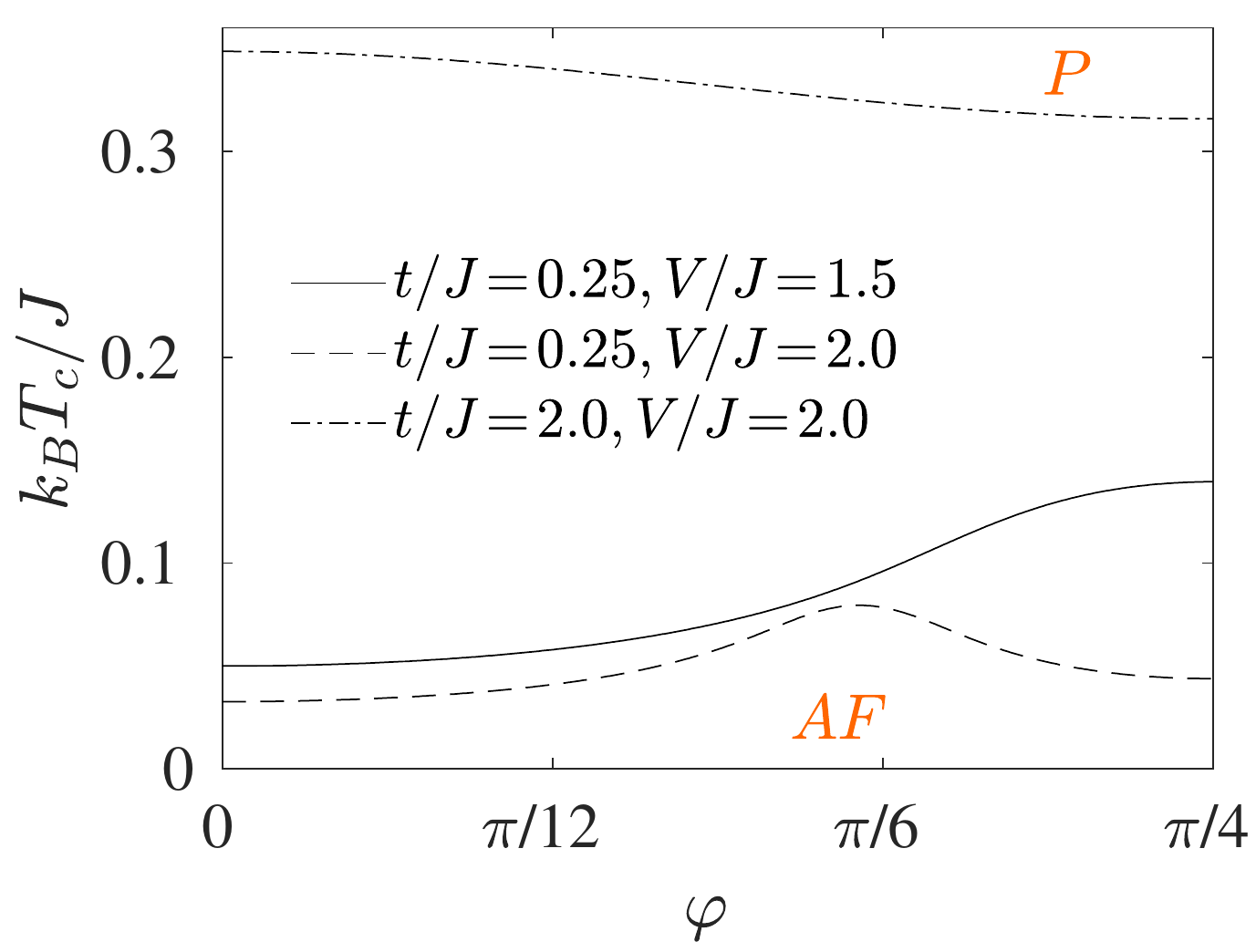}}
\caption{ The normalized critical temperature $k_BT_c/J$ as a function of the polar angle $\varphi$ for fixed values of the hopping term $t/J$ and the electric field $V/J$ directly illustrating the rotational magnetoelectric effect.}
\label{fig4}
\end{figure}
It turns out that the positions of detected maxima are very close to crossing points of the occurrence probabilities $a^{\delta}$ and $c^{\delta}$ defined through Eq.~(\ref{eq3b}). To illustrate the case, the occurrence probabilities are depicted in the lower panel of Fig.~\ref{fig3} for the particular case with $t/J\!=\!0.25$ and $\varphi\!=\!\pi/12$. For better clarity, the corresponding crossing points are visualized through vertical dashed lines also in the respective dependencies of the critical temperature (upper panel of Fig.~\ref{fig3}) with only a small deviation from the respective local maxima. It could be anticipated that this deviation originates from thermal fluctuations, which are completely neglected in the analysis of occurrence probabilities. Interestingly, thermally most stable antiferromagnetic state thus involves balanced occurrence probabilities for the antiferromagnetic and non-magnetic microstates of the mobile electrons. 

Last but not least, the critical temperature is plotted against the polar angle $\varphi$ in Fig.~\ref{fig4} to undoubtedly demonstrate the rotating magnetoelectric effect.  It is quite evident that the thermally most stable antiferromagnetic long-range order may emerge for a perfectly isotropic (e.g. $t/J\!=\!0.25$ and $V/J\!=\!1.5$), a strong (e.g. $t/J\!=\!2.0$ and $V/J\!=\!2.0$) or a weak (e.g. $t/J\!=\!0.25$ and $V/J\!=\!2.0$) anisotropic spatial orientation of the electric field depending on a relative size of the hopping term and the external field. It should be mentioned that the existence of a round maximum in the dependence critical temperature versus the polar angle $\varphi$ located in between the values 0 and $\pi/4$ is very rare and it is possible only for relatively weak hopping amplitudes $t/J\!<\!1.1$  on assumption that intersecting occurrence probabilities  $a^y$ and $c^y$ are disjoint with occurrence probabilities $a^x$ and $c^x$ forming their external envelope. 

\begin{figure}[t!]
{\includegraphics[width=.23\textwidth,height=2.9cm,trim=0.7cm 7.5cm 0.6cm 7.8cm, clip]{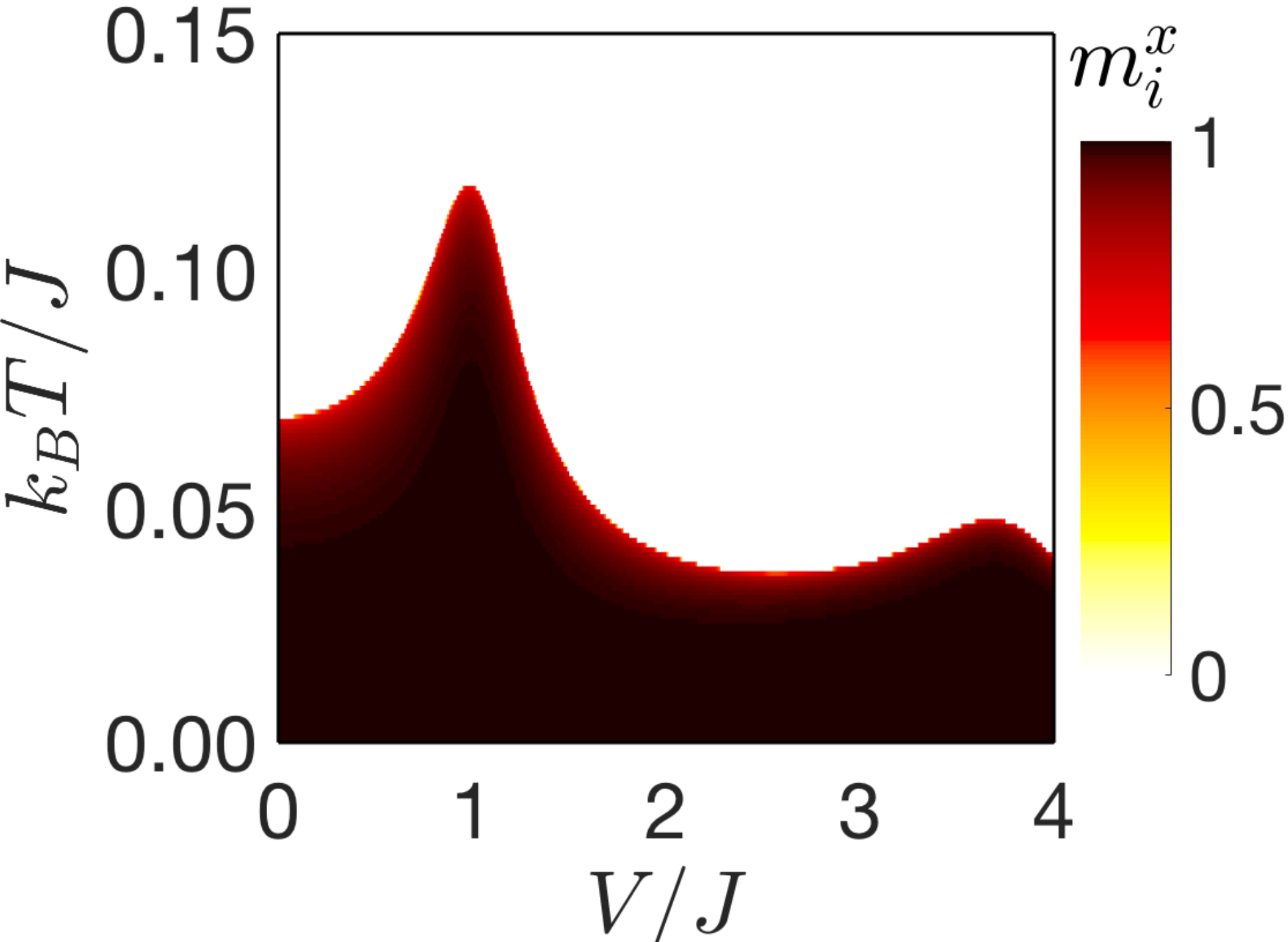}}
{\includegraphics[width=.23\textwidth,height=2.9cm,trim=0.7cm 7.5cm 0.6cm 7.8cm, clip]{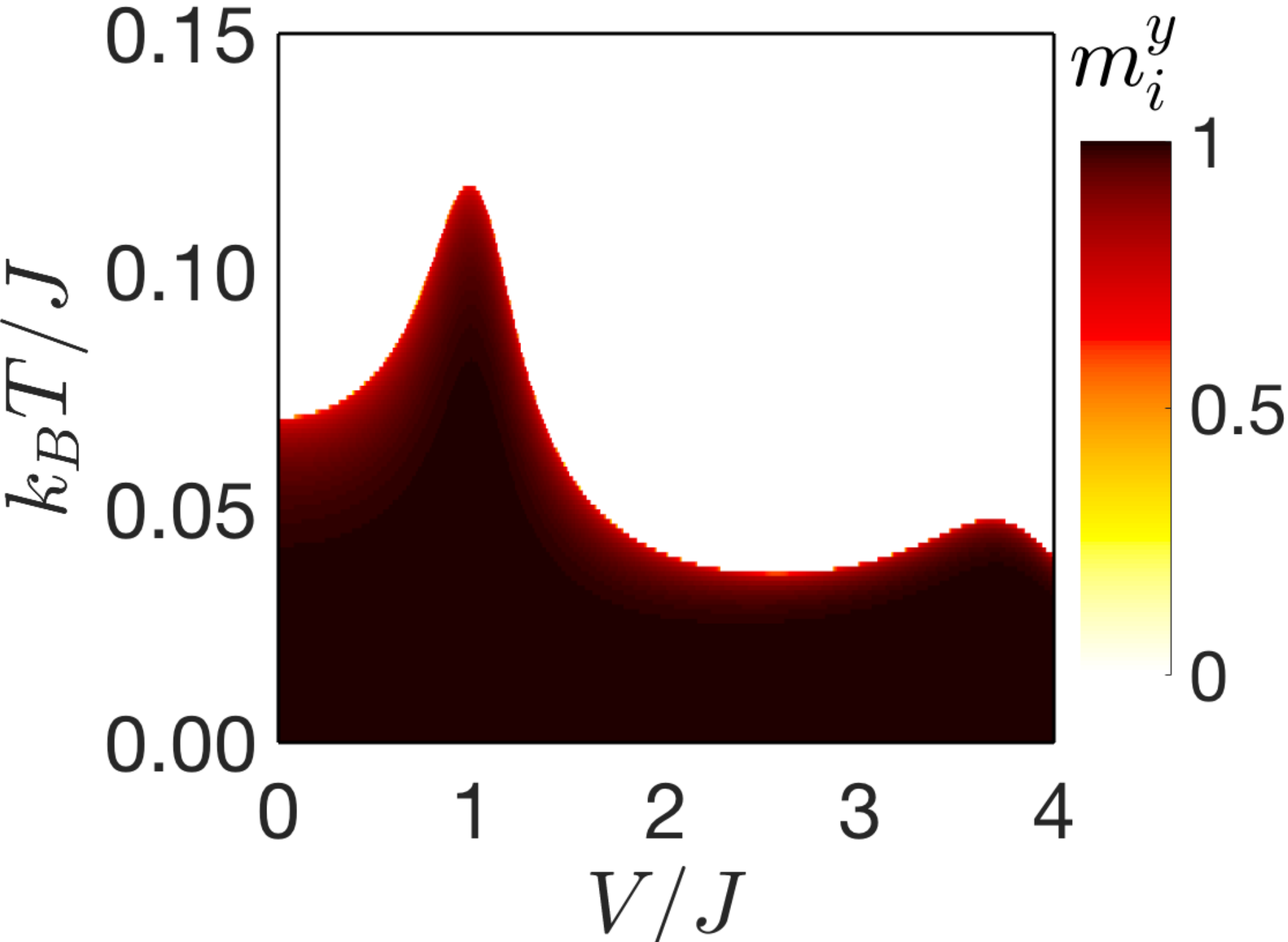}}
{\includegraphics[width=.23\textwidth,height=2.9cm,trim=0.7cm 7.5cm 0.6cm 7.8cm, clip]{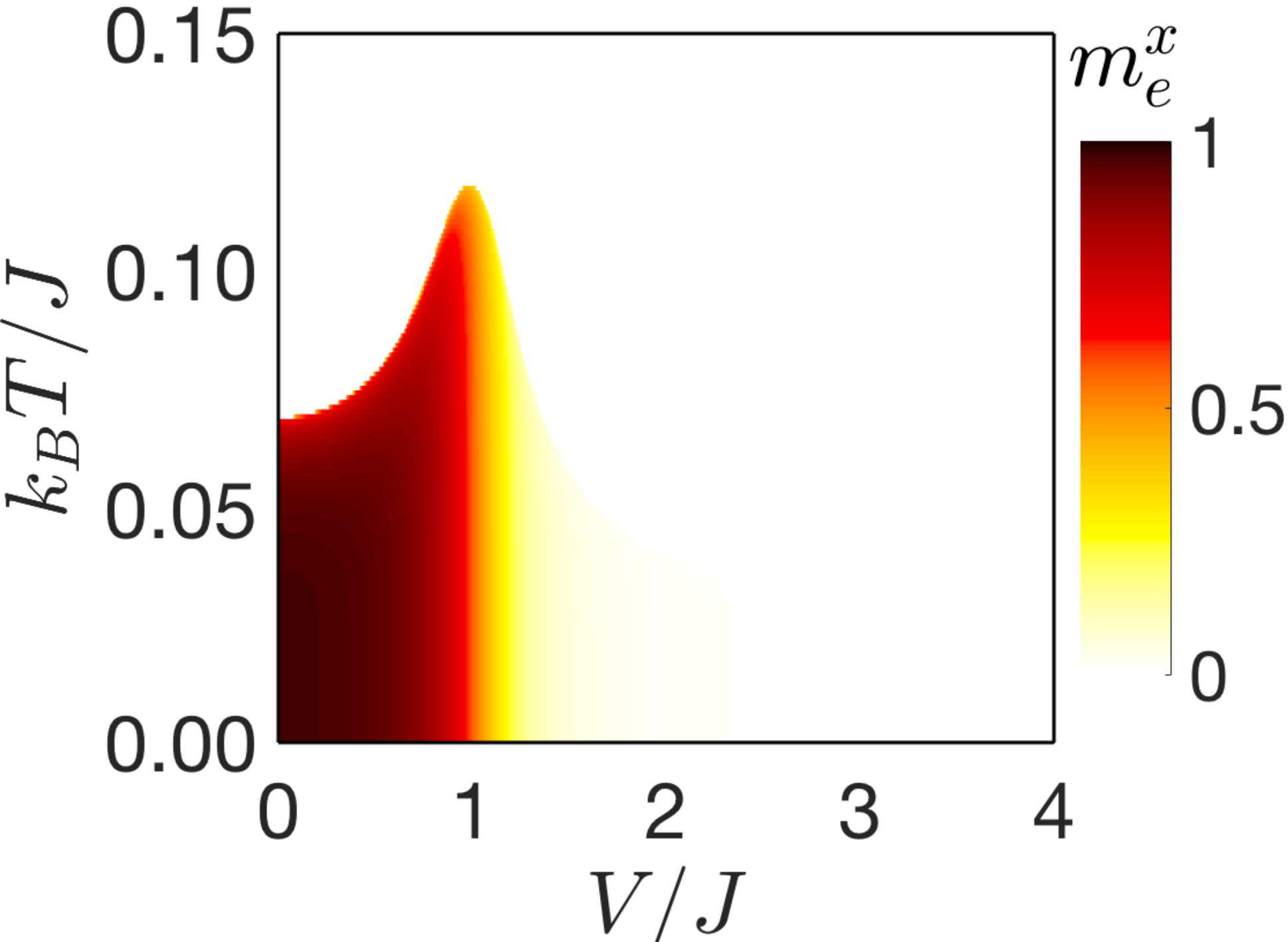}}
{\includegraphics[width=.23\textwidth,height=2.9cm,trim=0.7cm 7.5cm 0.6cm 7.8cm, clip]{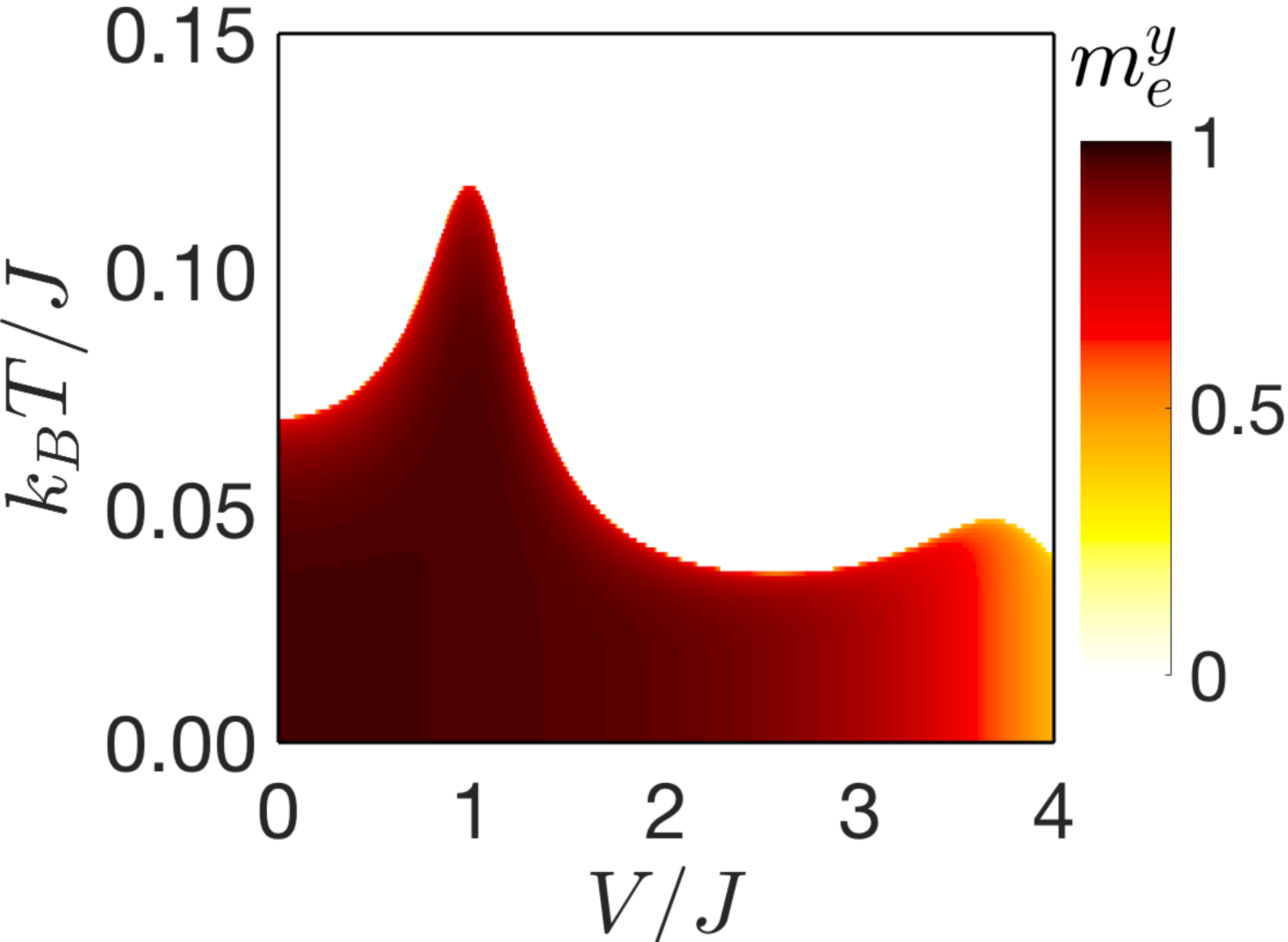}}
\caption{ Density plots of the staggered magnetization of the localized Ising spins ($m_i^{\delta}$) and the mobile electrons ($m_e^{\delta}$) in the electric-field size vs. temperature plane ($V/J$-$k_BT/J$) for the horizontal and vertical bonds ($\delta\!=\!x$ and $y$) by considering the fixed values of the hopping term $t/J\!=\!0.25$ and the polar angle $\varphi\!=\!\pi/12$.}
\label{fig5}
\end{figure}

\begin{figure}[h!]
{\includegraphics[width=.23\textwidth,height=2.9cm,trim=0.7cm 7.5cm 0.6cm 7.7cm, clip]{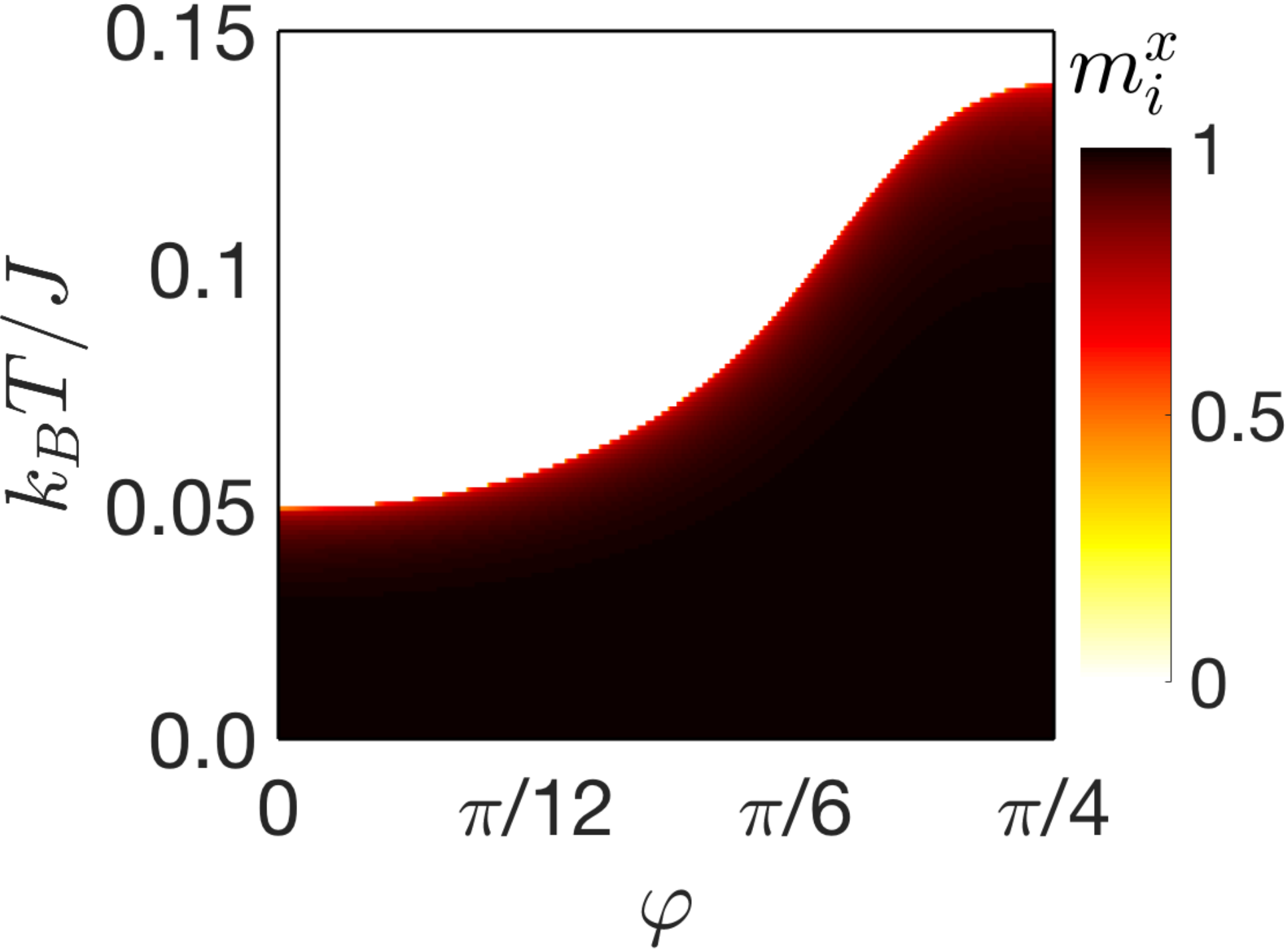}}
{\includegraphics[width=.23\textwidth,height=2.9cm,trim=0.7cm 7.5cm 0.6cm 7.7cm, clip]{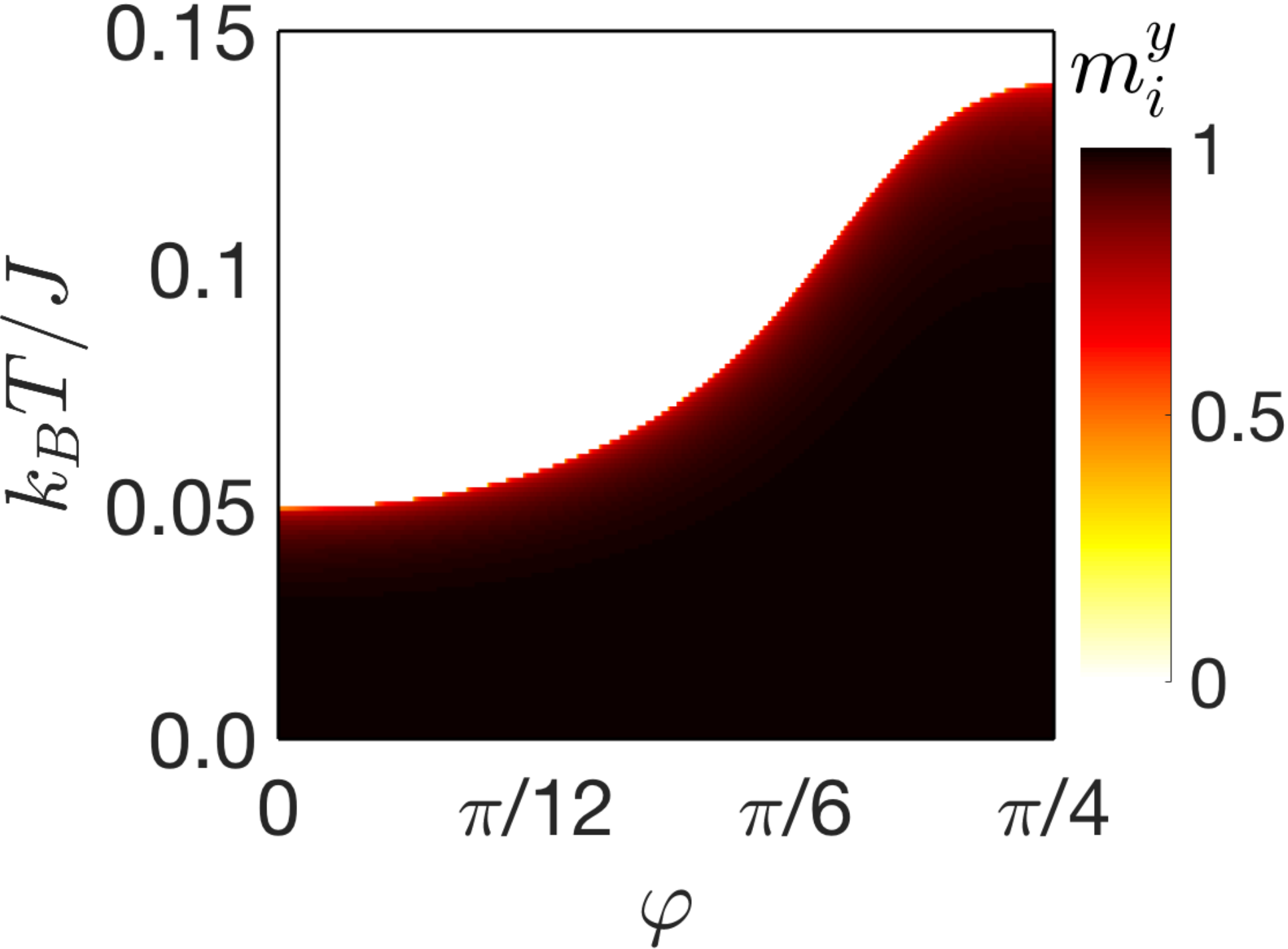}}
{\includegraphics[width=.23\textwidth,height=2.9cm,trim=0.7cm 7.5cm 0.6cm 7.7cm, clip]{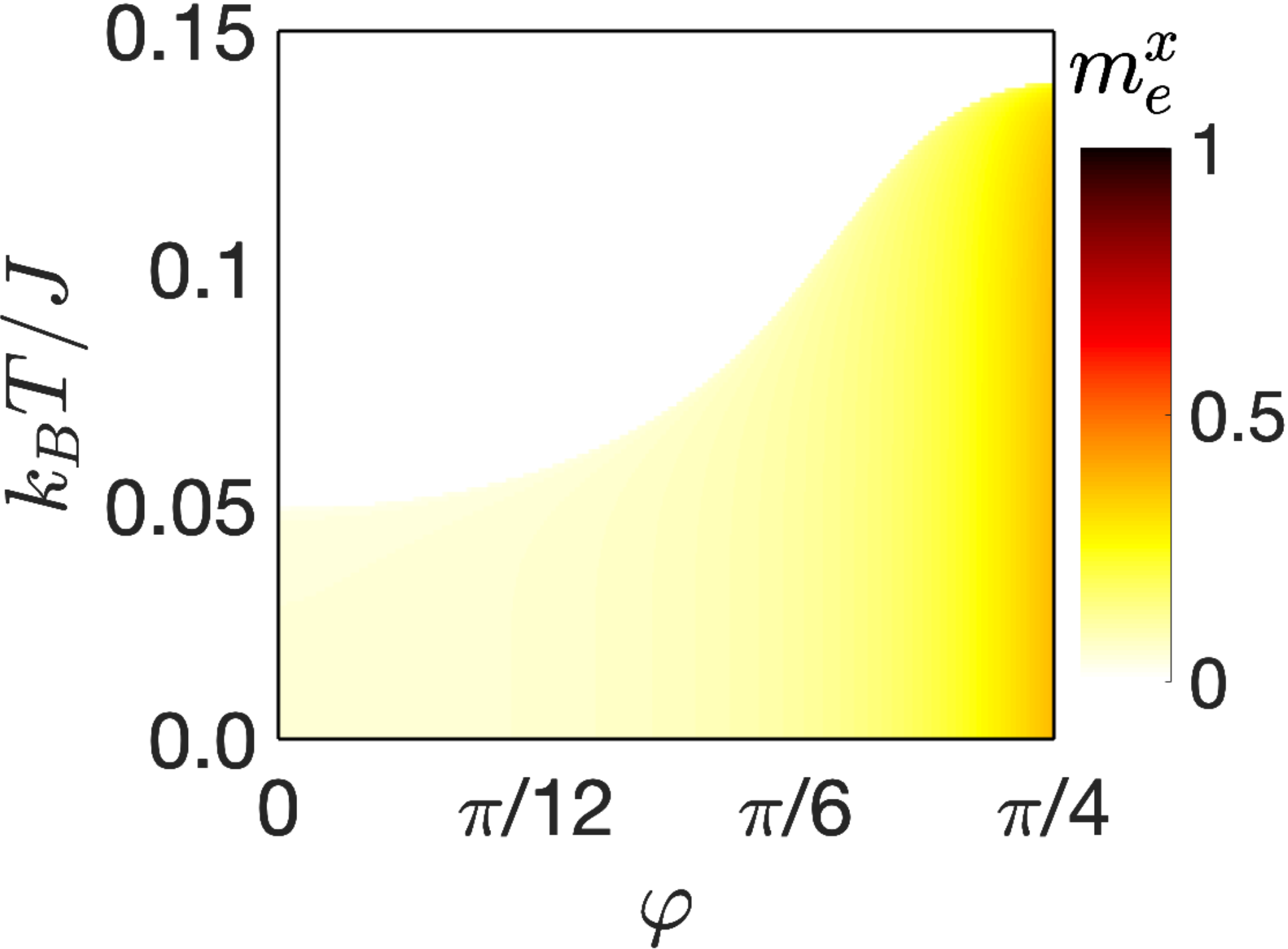}}
{\includegraphics[width=.23\textwidth,height=2.9cm,trim=0.7cm 7.5cm 0.6cm 7.7cm, clip]{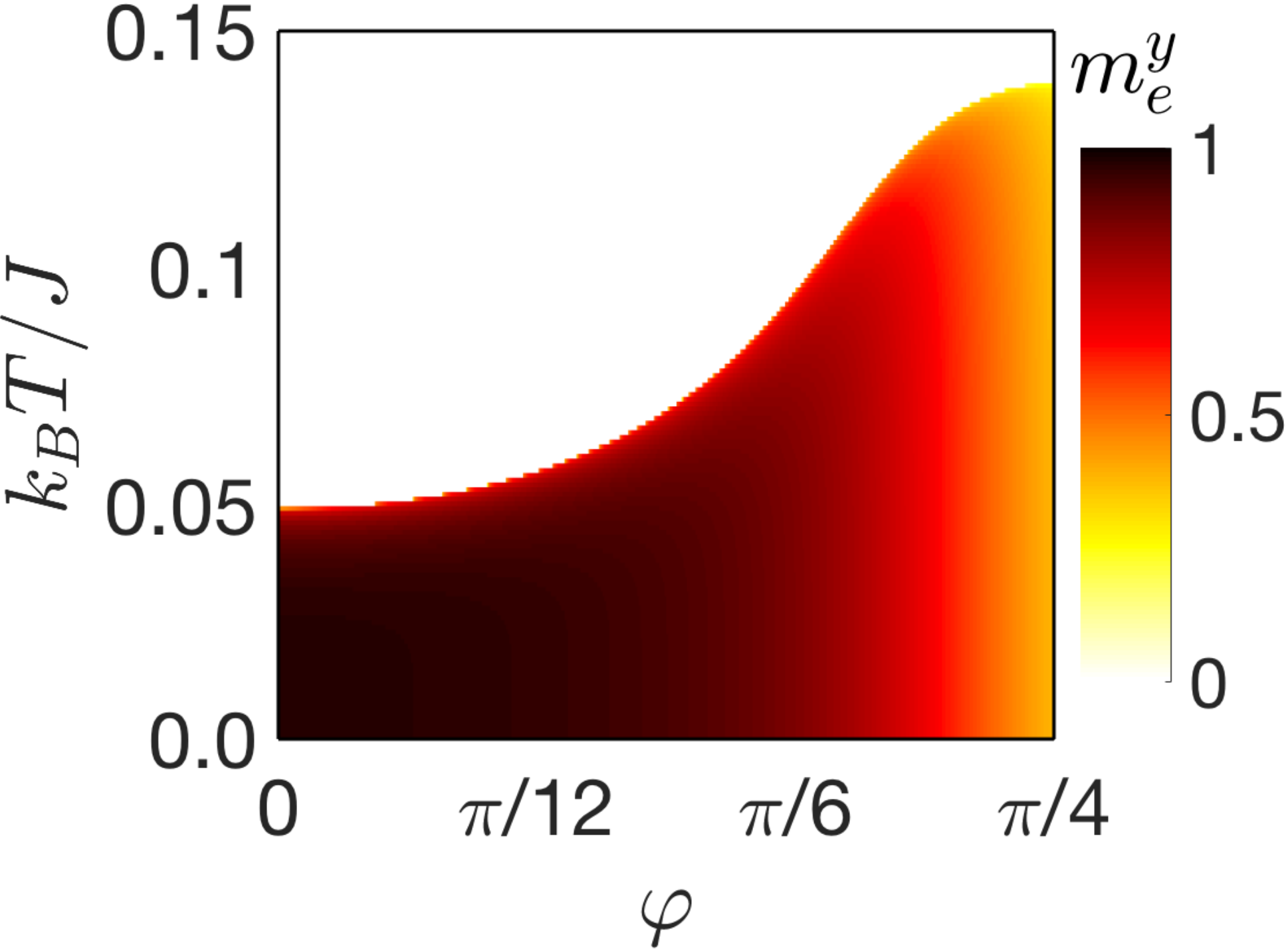}}
\caption{ Density plots of the staggered magnetization of the localized Ising spins ($m_i^{\delta}$) and the mobile electrons ($m_e^{\delta}$) in the electric-field orientation vs. temperature plane ($\varphi$-$k_BT/J$) for the horizontal and vertical bonds ($\delta\!=\!x$ and $y$) by considering the fixed values of the hopping term $t/J\!=\!0.25$ and the electric-field size $V/J\!=\!1.5$.}
\label{fig6}
\end{figure}

 Finally, the staggered magnetizations of the localized Ising spins and the mobile electrons are plotted in Fig.~\ref{fig5} and Fig.~\ref{fig6} in the electric-field size vs. temperature and the electric-field orientation vs. temperature planes, respectively. It is quite obvious from Fig.~\ref{fig5} and Fig.~\ref{fig6} that the staggered magnetization of the localized Ising spins $m_i^{\delta}$ behaves alike for both spatial directions $\delta\!=\!x$ and $y$ regardless of a relative size and orientation of the external electric field. This result is in a sharp contrast with the behavior of the staggered magnetization of the mobile electrons $m_e^{\delta}$, which might be quite different for two orthogonal directions $\delta\!=\!x$ and $y$ whenever the external electric field largely deviates from the isotropic case $\varphi\!=\!\pi/4$. As a matter of fact, it is quite clear from lower panels of Fig.~\ref{fig5} that the staggered magnetization at sufficiently high values of the electric field may even become zero in one spatial direction ($m_e^x\!=\!0$) while being nonzero ($m_e^y\!\neq\!0$) in the other one (see Fig.~\ref{fig5}). Hence, the investigated spin-electron model may exhibit a peculiar quasi-one-dimensional spontaneous antiferromagnetic long-range order, which is quite similar to that previously reported for the mixed-spin Ising model on a spatially anisotropic decorated square lattice \cite{str07}. 
In agreement with previous arguments, the critical temperature associated with a breakdown of the spontaneous staggered magnetization exhibits in Fig.~\ref{fig5} two round local maxima nearby crossing points of the occurrence probabilities of antiferromagnetic and non-magnetic microstates of the mobile electrons (c.f. Fig.~\ref{fig3}).  Beside this, Fig.~\ref{fig6} brings insight into how the rotational magnetoelectric effect influences the staggered magnetization of the mobile electrons in two orthogonal spatial directions. While the staggered magnetization $m_e^x$ in the first spatial direction is reinforced upon increasing of the polar angle $\varphi$, the staggered magnetization $m_e^y$ in the second spatial direction is contrarily lowered until they both converge to the same asymptotic value achieved for the polar angle $\varphi\!=\!\pi/4$.

\section{Conclusions}
\label{conclusions}
In the present Letter we have examined in detail  a conventional and rotating magnetoelectric effect in a half-filled spin-electron model on a doubly decorated square lattice.  Despite the fact that the ground state of the investigated spin-electron model is in the whole parameter space the unique quantum antiferromagnetic phase, it was found that thermal fluctuations give rise to a significant magnetoelectric effect affecting the critical temperature and spontaneous staggered magnetization serving as a stability criterion and order parameter of the quantum antiferromagnetic phase. Depending on a relative size of the applied electric field and hopping amplitude, the relative rotation of the electric field may stabilize or contrarily destroy the antiferromagnetic long-range ordering. A driving force, which determines the nature of a rotating magnetoelectric effect, lies  in a mutual competition between the occurrence probabilities of intrinsically antiferromagnetic and non-magnetic microstates of the mobile electrons. The rotating magnetoelectric effect generally becomes more significant at higher electric fields. In addition, an unexpected behaviour of the critical temperature has been observed at weak hopping amplitudes when the variation of the applied electric field may generate one or two consecutive round maxima. It has been deduced that these thermally most stabled antiferromagnetic states originate from a balanced character of the antiferromagnetic and non-magnetic microstates of the mobile electrons in one of two orthogonal directions. Moreover, the critical line with a single maximum is conditioned by the presumption that intersecting occurrence probabilities  $a^y$ and $c^y$ are disjoint with occurrence probabilities $a^x$ and $c^x$ forming their external envelope. Last but not least, it has been evidenced that a rotating magnetoelectric effect may favour the thermally most stabled antiferromagnetic state with a different internal structure, namely the homogeneous (isotropic) structure, or an inhomogeneous one with either strong or weak spatial anisotropy.
 
\begin{acknowledgments}
This work was financially supported by the grant of the Slovak Research and Development Agency provided under the contract No. APVV-16-0186. The financial support provided by the VEGA under the Grant No. VEGA 1/0043/16 is also gratefully acknowledged. H.C. also acknowledges support by the ERDF EU Grant under contract No. ITMS  26220120047.
\end{acknowledgments}

\end{document}